\documentclass[twocolumn,prb,superscriptaddress,showpacs,epsf]{revtex4}
\usepackage{graphicx}
\usepackage{epsf}
\usepackage{amsmath}
\usepackage[dvips]{color}
\begin{document}

\newcommand{\cmt}[1]{~~[{\textcolor{blue}{\it\small #1}}]~~}
\newcommand{\add}[1]{{\textcolor{red}{\it\small #1}}~~}
\newcommand{\rep}[1]{\add{Reply:}{\textcolor{red}{ \it\small #1}}}
\newcommand{\chk}[1]{{\textcolor{red}{#1}}}

\newcommand{\Li}{\mathcal L}
\newcommand{\Int}{{\rm int}}
\renewcommand{\otimes}{}

\title{Shot noise in electron transport through a double quantum dot: %
A master equation approach}
\date{\today}
\author{Shi-Hua Ouyang}
\affiliation{Department of Physics and Surface Physics Laboratory
(National Key Laboratory), Fudan University, Shanghai 200433, China}
\affiliation{Department of Applied Physics, Hong Kong Polytechnic
University, Hung Hom, Hong Kong, China}
\author{Chi-Hang Lam}
\affiliation{Department of Applied Physics, Hong Kong Polytechnic
University, Hung Hom, Hong Kong, China}
\author{J. Q. You}
\affiliation{Department of Physics and Surface Physics Laboratory
(National Key Laboratory), Fudan University, Shanghai 200433, China}

\begin{abstract}
  We study shot noise in tunneling current through a
  double quantum dot connected to two electric leads. We derive
  two master equations in the occupation-state basis and the eigenstate
  basis to describe the electron dynamics.
  The approach based on the occupation-state basis, despite widely
  used in many previous studies, is valid only when the interdot
  coupling strength is much smaller than the energy difference between
  the two dots. In contrast, the calculations using the eigenstate
  basis are valid for an arbitrary interdot coupling.
  We show that the master equation in the occupation-state basis
  includes only the low-order terms with respect to the interdot coupling
  compared with the more accurate master equation in the eigenstate
  basis.
  Using realistic model parameters, we demonstrate that the predicted
  currents and shot-noise properties from the two approaches are
  significantly different
  when the interdot coupling is not small.
  Furthermore,
  properties of the shot noise predicted using the eigenstate basis
  successfully reproduce qualitative features found in a recent
  experiment.
\end{abstract}

\pacs{72.70.+m, 73.63.Kv, 73.23.-b, 03.65.Yz} \maketitle

\section{Introduction}

Precise control of coherent coupling between quantum states is of
great importance in quantum information processing. Recent studies
show that artificial two-level systems designed using mesoscopic
circuits can be controlled in nanosecond time scales and can also
exhibit coherent oscillations between two quantum states (see, e.g.,
Refs.~\onlinecite{Nakamura99, You05, Koppens06}). A double quantum
dot (DQD) provides a useful system to explore coherent effects
because interdot hopping intrinsically couples states in two
different dots and is tunable via the gate
voltage.\cite{Petta04,Huttel05} A commonly used observable for
studying the effects of coherent coupling is the current through the
DQD.  Recently, shot-noise measurement has recently been
demonstrated as another useful tool to study the coherent
effects.\cite{Barthold07,Kiesslich07} Moreover, the shot-noise
properties have been predicted to be an indicator of the degree of
entanglement between electron states\cite{Neil07,Bodoky08} and they
are also related to the radiative decay properties of the
one-demensional quantum ring exciton.\cite{Chen05}

Shot noise, i.e., current fluctuations due to the discrete and
stochastic nature of electron transport, describes the correlation
between electrons transported successively through mesoscopic
systems, such as quantum dots (QDs) or molecular devices (for
reviews, see Refs.~\onlinecite{Blanter00,Nazarov03}). In classical
transport, the noise is typically Poissonian with a power density
$S=2e\langle I\rangle$, where $e$ is the unit charge and $\langle
I\rangle$ is the average current. However, either Coulomb
interaction or the Pauli's exclusion principle can induce a negative
correlation between successive transport events. This reduces the
noise power density so that $S<2e\langle I\rangle$ corresponding to
a sub-Poissonian noise.\cite{Chen92} In contrast, the interplay
between Coulomb interaction and the Pauli's exclusion principle can
also produce a positive correlation between the transport events,
i.e., $S>2e\langle I\rangle$. This corresponds to a super-Poissonian
noise. The Fano factor $F=S/2e\langle I\rangle$ is usually used to
characterize the shot noise, where $F=1$, $F>1$, or $F<1$,
respectively, corresponds to the Poissonian, super-Poissonian, or
sub-Poissonian noise. Many theoretical works show that
super-Poissonian noise of
electron\cite{Belzig05,Sanchez07,Sanchez08} or spin,
\cite{CottetEPL,Weymann08,Ouyang08,SanchezNJP} and positive cross
correlation between different spin states\cite{CottetPRL} in QDs can
be induced via dynamical channel blockade. Moreover, a
super-Poissonian noise in tunneling current caused by dynamical
channel blockade has been observed in a system consisting of two
electrostatically coupled QDs\cite{Zhang07} and also a single
QD\cite{Safonov03,Ozarchin07} in recent experiments.

In this work, we study the current and shot-noise in electrons
tunneling through a DQD. We apply two different approaches and
compare the results. First, we follow many previous investigations
(see, e.g., Refs.~\onlinecite{stoof96,Gurvitz96,Kiesslich07}) and
derive a master equation for the electron transport based on the
occupation-state basis of the DQD. We show that to arrive at this
master equation, one needs to assume that the interdot coupling
strength is much smaller than the energy difference between the two
dots. However, using realistic model parameters for the DQD, only
Poissonian or sub-poissonian shot noise is predicted, while
super-poissonian noise was also observed in a recent
experiment.\cite{Barthold07}

Alternatively, we derive a more generally applicable master equation
in the eigenstate basis of the DQD, which does not require the
assumption of a small interdot coupling and is hence valid for any
arbitrary interdot coupling strength.  The two master equations are
formally different in general and are identical only in the limiting
case when the interdot coupling is much smaller than the energy
difference between the two dots.  We show that for small interdot
coupling, the properties of the shot noise predicted by the two
master equation agree with each other as expected.  However, for
large interdot coupling, they are significantly different.  More
importantly, for typical model parameters, the shot noise deduced
using the master equation in the eigenstate basis exhibits rich
properties including Poissonian, sub-poissonian as well as
super-poissonian statistics in good agreement with recent
experimental observations (Ref.~\onlinecite{Barthold07}).
Furthermore, qualitative features of the current and the shot-noise
can easily be explained intuitively using the master equation in the
eigenstate basis.

The present paper is organized as follows. In Sec.~II, we introduce
the model for a DQD connected to two electric leads. A phonon bath
that affects the dynamics of the DQD is also considered. Two master
equations for the electron dynamics in the DQD are derived in both
the occupation-state basis and the eigenstate basis. In Sec. III and
Sec. IV, we study, respectively, the properties of the current
through the DQD and the associated shot noise. Results based on the
two master equations are compared. In Sec.~V, we discuss the
relation between the two master equations. A brief conclusion is
presented in Sec.~VI. Finally, Appendixes A and B give detailed
derivations of the master equations in both the occupation-state
basis and the eigenstate basis.

\section{Time evolution of the reduced density matrix of a double quantum dot}
\begin{figure}[tbp]
\includegraphics[width=3.0in,
bbllx=200,bblly=103,bburx=563,bbury=538]{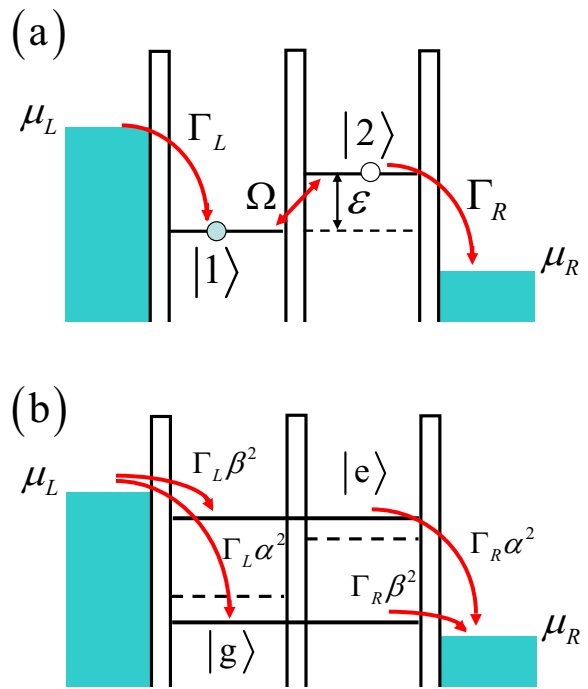} \caption{(Color
online)~Schematic diagram of an electron transported through a DQD
connected to two electric leads via tunneling barriers. (a)
Considering DQD electron states in the occupation-state basis, the
electron tunnels sequentially from the left lead to the right lead
via first the left dot and then the right dot.  (b) Considering the
eigenstate basis, the electron is transport via either the
ground-state channel or and the excited-state channel. The effective
tunneling rates from the left lead to the ground state and the
excited state are $\Gamma_L\alpha^2$ and $\Gamma_L\beta^2$, while
those from the ground state and the excited state to the right lead
are $\Gamma_R\beta^2$ and $\Gamma_R\alpha^2$. \label{fig1}}
\end{figure}
The schematic diagram of a DQD connected to two electrodes by
tunneling barriers is shown in Fig.~\ref{fig1}. The voltage across
the DQD is biased so that the chemical potential of the left
electrode $\mu_L$ is higher than that of the right electrode
$\mu_R$. Thus, an electron can tunnel from the left electrode to the
right one via the DQD. We assume that the DQD is in the Coulomb
regime (with both strong intra- and interdot Coulomb repulsions), so
that at most a single electron is allowed in the DQD. In the
occupation representation, the electron basis states are the vacuum
state $|0\rangle$, the state with one electron in the left dot
$|1\rangle$, and the state with one electron in the right dot
$|2\rangle$.

The Hamiltonian of the whole system reads (taking
$\hbar=1$)
\begin{equation}
H_{\rm{tot}}\!=\!H_{\rm{leads}}+H_{\rm{DQD}}+H_{\rm{T}}+H_{\rm{ph}}+H_{\rm{ep}}.\label{H-total}
\end{equation}
The first two terms $H_{\rm{leads}}$ and $H_{\rm{DQD}}$ are
respectively the Hamiltonians of the two electrodes and the DQD, and
are given by
\begin{eqnarray}
H_{\rm{leads}}&=&\sum_{\alpha k}\omega_{\alpha k}c_{\alpha k}^{\dagger}%
c_{\alpha k},\\
H_{\rm{DQD}}&=&\frac{\varepsilon}{2}\sigma_z+\Omega\sigma_x,
\label{H-DQD}
\end{eqnarray}
where $c_{\alpha k\sigma}^{\dagger}$ ($c_{\alpha k\sigma}$) is the
creation (annihilation) operator of an electron with momentum $k$ in
the electrode $\alpha$ $(\alpha=l,r)$;
$\sigma_z=a_2^{\dagger}a_2-a_1^{\dagger}a_1$ and
$\sigma_x=a_2^{\dagger}a_1+~a_1^{\dagger}a_2$ are the Pauli
matrices, with $a_1^{\dagger}$ ($a_2^{\dagger}$) being the electron
creation operator in the left (right) dot of the DQD. In
Eq.~(\ref{H-DQD}), the first term $\varepsilon\sigma_z/2$, with
$\varepsilon=\varepsilon_2-\varepsilon_1$ denoting the energy
difference between the two dots, gives the Hamiltonian of the two
uncoupled quantum dots, while the second term $\Omega\sigma_x$
characterizes the interdot hopping. The tunneling coupling between
the DQD and the electrodes is described by
\begin{eqnarray}
H_{\rm{T}}=\sum_{k}(\Omega_{lk}\;a_{1}^{\dagger}\,c_{lk}+\Omega_{rk}\;
a_{2}^{\dagger}\,\Upsilon_r\,c_{rk}+\rm{H.c.}),\label{HT-O}
\end{eqnarray}
where $\Omega_{lk(rk)}$ is the tunneling strength between the QD and
the left (right) electrode. The operators $\Upsilon_r$
($\Upsilon_r^{\dagger}$) decreases  (increases) the number of
electrons having tunneled into the right lead (via the barrier
between the DQD and the right lead).\cite{Doiron07} These counting
operators allow one to keep track of the tunneling process during
the evolution of the DQD. Below we focus on current and shot noise
in electron tunneling through the right tunneling barrier, so only
the related counting operators ($\Upsilon_r$ and
$\Upsilon_r^{\dagger}$) are introduced in the tunneling Hamiltonian
(\ref{HT-O}).

Also, we consider the effects of the phonon-bath environment on the
evolution of the DQD. The Hamiltonian of this phonon bath is
\begin{eqnarray}
H_{\rm{ph}}=\sum_{q}\omega_{q}b_{q}^{\dagger}%
b_{q},
\end{eqnarray}
with $b_q^{\dagger}$ ($b_q$) creating (annihilating) a phonon with
frequency $\omega_q$. The electron-phonon interaction is given by
\begin{eqnarray}
H_{\rm{ep}}=
\sigma_x\sum_{q}\lambda_q(b_q^{\dagger}+b_q),\label{Hsb-O}
\end{eqnarray}
where $\lambda_q$ is the electron-phonon coupling strength.

The evolution of the whole system is described by the von Neumann
equation for the density matrix $\rho_R$ of the whole system:
\begin{eqnarray}
\dot{\rho}_{R}(t) = -i[\,H_{\rm{tot}},~\rho_{R}(t)\,].\label{van
Neumann}
\end{eqnarray}
Here we are interested in the time evolution of the DQD and treat
both the electric leads and the phonon bath as the total outside
environment. We will hence derive the master equation of the reduced
density matrix $\rho_{d}$ of the DQD: $\rho_{d}\equiv\,{\rm
Tr}_E\{\rho_R\}$, where ${\rm Tr}_{E}\{\cdots\}$ denotes the trace
over the degrees of freedom of both the electric leads and the
phonon bath. In our calculations, we adopt
the interaction picture based on the
free Hamiltonian
\begin{equation}
H_0= H_{\rm{leads}}+H_{\rm{ph}}+H_{\rm{DQD}},\label{Free-H-1}
\end{equation}
and the interaction Hamiltonian becomes
\begin{eqnarray}
H_{\rm int}(t)&=& H_{\rm T}(t)+ H_{\rm ep}(t), \label{H-int}
\end{eqnarray}
where operators in the interaction and the Schrodinger pictures are
related by $A(t) = e^{iH_0t}Ae^{-iH_0t}$ for any operator $A$.

After tracing over the degrees of freedom of both the electrodes and
the phonon bath, one obtains the master equation for the reduced
density matrix $\rho_d^I$ of the DQD in the interaction picture
as\cite{Blum}
\begin{eqnarray}
\dot{\rho}^I_{ d}(t)&=&-i\,{\rm Tr}_{E}[\,H_{\rm int}(t),~\rho_{
d}(0)\rho_E(0)\,]\nonumber\\
&&-{\rm Tr}_{E}\int_{0}^t dt'[\,H_{\rm int}(t),\,[\,H_{\rm
int}(t'),\,
\rho^I_d(t')\otimes\rho_{E}(0)]\;],\nonumber\\&&\label{ME-1}
\end{eqnarray}
where $\rho_{E}$
is the density matrix of the outside environment.
Because the trace of a single unpaired creation or annihilation
operator over the lead or the phonon bath is zero, e.g., ${\rm
Tr}_{E}\{c_{lk}\rho_E\}=0$, the first term in Eq.~(\ref{ME-1})
vanishes.
Within the
Born-Markov approximation, we have
\begin{eqnarray}
\dot{\rho}^I_{ d}(t) = -{\rm Tr}_E\int_{0}^t dt'[\,H_{\rm
int}(t),\,[\,H_{\rm int}(t'),\,
\rho^I_d(t)\otimes\rho_E(0)]\;],\nonumber\\&&\label{ME-2}
\end{eqnarray}
Here the Born approximation amounts to the use of the second-order
perturbation theory with respect to the interaction Hamiltonian
$H_{\rm int}$, while the Markov approximation assumes that the
correlation times of the outside environment (both the electric
leads and the phonon bath) are much shorter than the typical
quantum-state evolution time of the DQD.

Since the left lead, the right lead and the phonon bath are
completely independent of each other, the density matrix of the
outside environment $\rho_{E}$ can be written as a tensor product of
density matrices that describe the subsystems, i.e.,
$\rho_{E}=\rho_{L}\otimes\rho_{R}\otimes\rho_{\rm ph}$, where
$\rho_L$, $\rho_R$ and $\rho_{\rm ph}$ are, respectively, the
density matrices of the left lead, the right lead and the phonon
bath. Therefore, the trace of the integrand in Eq.~(\ref{ME-2}) can
be expressed as:
\begin{eqnarray}
&&{\rm Tr}_{E}[\,H_{\rm int}(t),\;[\,H_{\rm int}(t'),\,
\rho^I_d(t)\otimes\rho_E]\;]\nonumber\\
\!&\!=\!&\!\sum_{\alpha=l,r}{\rm Tr_{\alpha}}[\,H_{\rm
T}(t),\,[\,H_{\rm T}(t'), \rho^I_d(t)\otimes\rho_{\rm \alpha}]\;]
\nonumber\\&& +\,{\rm Tr_{ph}}\,[\,H_{\rm ep}(t),\,[\,H_{\rm
ep}(t'),\; \rho^I_d(t)\otimes\rho_{\rm ph}]\;].\label{DIVIDE}
\end{eqnarray}
Equation~(\ref{ME-2}) can thus be written as a
sum of two corresponding parts:
\begin{eqnarray}
\dot{\rho}_d^I(t)\!&\!=\!&\!\mathcal{L}_{\rm
T}\rho_d^I(t)+\mathcal{L}_{\rm ph}\rho_d^I(t).\label{A-ME-O}
\end{eqnarray}
Here,
the dissipative part due to the electric leads is given by
\begin{eqnarray}
\mathcal{L}_{\rm T}\rho_d^I(t)\!&\!=\!&\!-\sum_{\alpha=l,r}{\rm
Tr}_{\alpha}\int\limits_0^{t}dt'\big[\,H_{\rm T}(t)H_{\rm
T}(t')\rho_d^I(t)\,\rho_{\rm leads}\nonumber\\&&
-H_{\rm T}(t)\rho_d^I(t)\,\rho_{\rm leads}H_{\rm T}(t') +{\rm
H.c.}\,\big],\label{Tunneling}
\end{eqnarray}
where $\rho_{\rm leads}=\rho_L\otimes\rho_R$ is the density matrix
of the two electric leads. The dissipative part caused by the phonon
bath is
\begin{eqnarray}
\mathcal{L}_{\rm ph}\rho_d^I(t)\!&\!=\!&\!- {\rm Tr}_{\rm
ph}\int\limits_0^{t}dt'\big[\,H_{\rm ep}(t)H_{\rm
ep}(t')\rho_d^I(t)\,\rho_{\rm ph}\nonumber\\&&
-H_{\rm ep}(t)\rho_d^I(t)\,\rho_{\rm ph}H_{\rm ep}(t')+{\rm
H.c.}\,\big].\label{Electron-Phonon}
\end{eqnarray}

From Eqs.~(\ref{A-ME-O})--(\ref{Electron-Phonon}), one can derive
the master equation for the $n$-resolved reduced density matrix
$\rho^{(n)}_d(t)$ of the DQD, where $\rho^{(n)}_d(t)\equiv\langle
n|\rho_d(t)|n\rangle$ and $n$ is the number of electrons that have
arrived at the right lead at time $t$. Below we derive two versions
of the master equation in both the occupation-state basis and the
eigenstate basis and then use them independently to study the
current and shot-noise properties of the DQD.

\subsection{Master equation in the occupation-state basis}

The master equation of the DQD in the occupation-state basis was
previously used to study the current
properties\cite{stoof96,Gurvitz96} and shot-noise
properties\cite{Kiesslich07,Weymann08,SanchezNJP,Sanchez08} of
electrons tunneling through the DQD. The occupation-state basis is
defined by the states $|0\rangle$, $|1\rangle$, and $|2\rangle$,
which correspond to the states of an empty DQD, one electron in the
left dot, and one electron in the right dot, respectively. In the
interaction picture defined by the free Hamiltonian $H_0$ in
Eq.~(\ref{Free-H-1}), the unperturbed evolution operator
$U_0(t)=e^{iH_0t}$ is difficult to calculate in the occupation-state
basis in the presence of interdot coupling. One hence split $H_0$
into two parts:
\begin{equation}
  H_0= H_{1}+H_{\Omega},\label{H-0}
\end{equation}
where
\begin{eqnarray}
  H_1 &=& H_{\rm{leads}}+H_{\rm{ph}}+
  \frac{\varepsilon}{2}\sigma_z, \label{H-1}\\
  H_\Omega &=& \Omega\sigma_x.\label{H-interdot}
\end{eqnarray}
Following previously works\cite{Goan01,Korotkov99} on deriving
master equations in the occupation-state basis, we assume that the
interdot couping $\Omega$ is small and satisfies
$\Omega\ll|\varepsilon_2-\varepsilon_1|$, we have $H_0 \simeq H_1$.
The evolution operator can then be approximated as
\begin{equation}
  \label{U0}
  e^{iH_0t} \simeq e^{i H_1t}.
\end{equation}
With this approximation, one easily obtains
\begin{eqnarray}
H_{\rm T}(t)
\!&\!\approx\!&\!\sum_{k}[\,\Omega_{lk}\;a_{1}^{\dagger}\,c_{lk}e^{-i(\omega_{lk}-\omega_1)t}\nonumber\\
&&+\;\Omega_{rk}\,
\Upsilon_r\,a_{2}^{\dagger}\,c_{rk}e^{-i(\omega_{rk}-\omega_2)t}+\rm{H.c.}\,],\label{H-T(t)}
\end{eqnarray}
and
\begin{eqnarray}
H_{\rm ep}(t)\!&\!\approx\!&\!(\sigma_+e^{i\varepsilon
t}+\sigma_-e^{-i\varepsilon
t})\sum_{q}\lambda_q(b_q^{\dagger}e^{i\omega_q
t}+b_qe^{-i\omega_qt}),\nonumber\\&&\label{H-sb(t)}
\end{eqnarray}
where $\omega_{1,2}\equiv\mp\varepsilon/2$. Here
$\sigma_+\!=\!a_2^{\dagger}a_1$ and
$\sigma_-\!=\!\sigma_{+}^{\dagger}$ are the raising and lowing
operators in the occupation-state basis.

We now consider the nonequilibrium case with a large bias voltage
across the DQD, so that all energy levels of the DQD lie within the
bias window, as shown in Figs.~\ref{fig1}(a) and \ref{fig1}(b).
Substituting Eqs.~(\ref{H-T(t)}) and (\ref{H-sb(t)}) into
Eqs.~(\ref{A-ME-O})--(\ref{Electron-Phonon}), taking the trace over
the degrees of freedom of both the two electrodes and the phonon
bath, and converting the obtained equation to the Schr\"{o}dinger
picture, one arrives at the master equation in the occupation-state
basis for a {\it weak} interdot coupling
$\Omega\ll|\varepsilon_2-\varepsilon_1|$ (see Appendix A):
\begin{eqnarray}
\dot{\rho}_d(t)\!&\!=\!&\!-i[\,H_{\rm DQD},\,\rho_d]+
\frac{\Gamma_L}{2}\mathcal{D}[a_1^{\dagger}]\rho_d +
\frac{\Gamma_R}{2}\mathcal{D}[\Upsilon_r^{\dagger}a_2]\rho_d\nonumber\\
&&+\frac{\gamma_1}{2}\mathcal{D}[\sigma_-]\rho_d+\frac{\gamma_2}{2}\mathcal{D}[\sigma_+]\rho_d,\label{ME-O}
\end{eqnarray}
where $\Gamma_{L(R)}=2\pi\rho_{lk(rk)}\Omega_{lk(rk)}^2$ is the
electron tunneling rate through the left (right) tunneling barrier.
Here, the electron density of states $\rho_{\alpha k}$ at lead
$\alpha$ ($\alpha=l,r$) and the tunneling strength $\Omega_{\alpha
k}$ are assumed to be energy-independent. The notation $\mathcal{D}$
acting on any operator $A$ is defined as
\begin{equation}
\mathcal{D}[A]\rho=2A\rho A^{\dagger}-[A^{\dagger}A\rho+\rho
A^{\dagger}A].
\end{equation}
The dissipation rates induced by the electron-phonon interaction are
\begin{eqnarray}
\gamma_1 &=& 2\pi\big\{ {{J}\left(\varepsilon \right) \left[{n\left(
\varepsilon  \right) + 1} \right] +
{J}\left( { - \varepsilon } \right)n\left( { - \varepsilon } \right)} \big\},\nonumber\\
\gamma _2 &=& 2\pi\big\{ {{J}\left( { - \varepsilon } \right)\left[
{n\left( { - \varepsilon } \right) + 1} \right] + {J}\left(
\varepsilon  \right)n\left( \varepsilon
\right)}\big\},\label{dissipation rate-O}
\end{eqnarray}
where
\begin{equation}
{J}(\omega)=\sum\limits_q { \lambda _q^2\;\delta \left( {\omega -
\omega _q } \right)},
\end{equation}
is the bath spectral density and $n\left(\omega\right)\!=\![{{\exp
\left( {\omega/k_B T} \right) - 1}}]^{-1}$
is the average phonon number at temperature $T$. Using
Eq.~(\ref{ME-O}) and the relations:\cite{Doiron07}
\begin{eqnarray}
\langle
n|\Upsilon_r^{\dagger}\Upsilon_r\rho_d|n\rangle&=&\rho_d^{(n)},\nonumber\\
\langle
n|\Upsilon_r\Upsilon_r^{\dagger}\rho_d|n\rangle&=&\rho_d^{(n)},\nonumber\\
\langle
n|\Upsilon_r^{\dagger}\rho_d\Upsilon_r|n\rangle&=&\rho_d^{(n-1)},\nonumber\\
\langle
n|\Upsilon_r\rho_d\Upsilon_r^{\dagger}|n\rangle&=&\rho_d^{(n+1)},\label{N-relation}
\end{eqnarray}
one obtains the equation of motion for each density matrix element:
\begin{eqnarray}
\dot \rho_{\rm 00}^{(n)}(t)\!&\!=\!&\!-\Gamma_L\rho_{\rm 00}^{(n)}+\Gamma_R\rho_{22}^{(n-1)},\nonumber\\
\dot \rho _{11}^{(n)} \left( t \right)\!&\!=\!&\! \Gamma _L \rho
_{00}^{(n)} + i\,\Omega \left( {\rho _{12}^{(n)}  - \rho _{21}^{(n)}
} \right)
+ \gamma _1 \rho _{22}^{(n)}  - \gamma _2 \rho _{11}^{(n)}, \nonumber\\
\dot \rho _{22}^{(n)} \left( t \right)\!&\!=\!&\!- \Gamma _R \rho
_{22}^{(n)} - i\,\Omega \left( {\rho _{12}^{(n)}  - \rho _{21}^{(n)}
} \right)
- \gamma _1 \rho _{22}^{(n)}  + \gamma _2 \rho _{11}^{(n)},  \nonumber\\
\dot\rho_{12}^{(n)}(t)\!&\!=\!&\!i\varepsilon \rho _{12}^{(n)} +
i\,\Omega \left( {\rho _{11}^{(n)}  - \rho _{22}^{(n)} } \right) -
\frac{{\Gamma _R  + \gamma _1  + \gamma _2 }}{2}\rho _{12}^{(n)}.
\nonumber\\\label{EOM-elements-O}
\end{eqnarray}
Then, the $i$th diagonal matrix element
$\rho_{ii}=\sum_n\rho_{ii}^{(n)}$ ($i=0,1$, or $2$) gives the
occupation probability of the state $|i\rangle$. The off-diagonal
matrix element $\rho_{12}(t)=\sum_n\rho_{12}^{(n)}$ describes the
coherence between states $|1\rangle$ and $|2\rangle$, and
$\rho_{21}(t)=\rho_{12}^{\ast}(t)$. This master equation was used in
many previous studies, e.g., Refs.~\onlinecite{stoof96} and
\onlinecite{Gurvitz96}.

The physical meaning of the master equation can be understood as
follows. Take the equation for $\rho_{11}^{(n)}$ in
Eq.~(\ref{EOM-elements-O}) for example. The first term on the
right-hand side describes the process of an electron tunneling from
the left lead to the left dot with rate $\Gamma_L$. The second term
represents the coherent coupling between states $|1\rangle$ and
$|2\rangle$ due to the interdot coupling. The third term describes
the phonon-induced relaxation process from state $|2\rangle$ to
$|1\rangle$ with rate $\gamma_1$. Finally, the fourth term describes
the inverse process with rate $\gamma_2$.

\subsection{Master equation in the eigenstate basis}

As explained above, the master equation in the occupation-state basis
is only valid for a weak interdot coupling. To extend the results to
an arbitrary interdot coupling $\Omega$, we now derive the master
equation in the eigenstate basis of the DQD. The result is valid for
any arbitrary interdot coupling strength.

Diagonalizing the Hamiltonian of the DQD [Eq.~(\ref{H-DQD})], one
has
\begin{equation}
\label{diagonalized} H_{\rm
DQD}=\frac{\Omega_0}{2}\,(\,|e\rangle\langle e|-|g\rangle\langle
g|\,)=\frac{\Omega_0}{2}\,\sigma_z^{(e)},
\end{equation}
where $\Omega_0=\sqrt{\varepsilon^2+4\,\Omega^2}$ is the energy
splitting of the two eigenstates of the DQD given by
\begin{eqnarray}
|e\rangle&=&\sin\frac{\theta}{2}\,|1\rangle+\cos{\frac{\theta}{2}}\,|2\rangle,
\nonumber\\
|g\rangle&=&\cos\frac{\theta}{2}\,|1\rangle-\sin{\frac{\theta}{2}}\,|2\rangle,
\label{basis-transform}
\end{eqnarray}
with $\tan\theta=2\Omega/\varepsilon$. The eigenstates and the
occupation states are related by
\begin{eqnarray}
\left| 1 \right\rangle  &=& \cos \frac{\theta}{2}\,\left| g
\right\rangle  +\, \sin \frac{\theta }{2}\,\left|e\right\rangle
,\nonumber\\
\left| 2 \right\rangle  &=&  - \sin \frac{\theta }{2}\,\left| g
\right\rangle + \cos \frac{\theta }{2}\,\left| e
\right\rangle.\label{relation}
\end{eqnarray}
With these relations, the tunneling Hamiltonian [Eq.~(\ref{HT-O})]
and the electron-phonon interaction [Eq.~(\ref{Hsb-O})] can be
written, in the eigenstate basis, as
\begin{eqnarray}
H_{\rm{T}}\!&\!=\!&\!\sum_{k}\bigg[\Omega_{lk}\;{\big(\,{\cos
\frac{\theta }{2}\,a_g^{\dagger}+\, \sin \frac{\theta
}{2}\,a_e^{\dagger}}\,\big)}
\,c_{lk}\nonumber\\
&&+\,\Omega_{rk}\; {\big(\, { - \sin \frac{\theta
}{2}\,a_g^{\dagger}+ \,\cos \frac{\theta
}{2}\,a_e^{\dagger}}\,\big)}
\,\Upsilon_r\,c_{rk}+\rm{H.c.}\bigg],\nonumber\\
H_{\rm ep}\!&\!=\!&\!\left[ {\sin \theta \,\,\sigma_z^{(e)}+ \cos
\theta \,\,\sigma_x^{(e)}} \right]\sum\limits_q {\lambda _q \left(
{b_q^{\dagger} + b_q } \right)}.\nonumber\\
\end{eqnarray}
In the interaction picture
based on the free Hamiltonian $H_0$ given by
Eq. (\ref{Free-H-1}), they become
\begin{eqnarray}
H_{\rm{T}}(t)\!&\!=\!&\!\sum_{k}\big\{\Omega_{lk}{\big(\,{\cos
\frac{\theta }{2}\,a_g^{\dagger}e^{i\omega_{g}t}+\sin \frac{\theta
}{2}\,a_e^{\dagger}e^{i\omega_{e}t}}\,\big)} c_{lk}
\nonumber\\
&&e^{-i\omega_{lk}t}+\Omega_{rk}{\big[{-
\sin\frac{\theta}{2}a_g^{\dagger}e^{i\omega_{g}t}+ \,\cos
\frac{\theta }{2}\,a_e^{\dagger}}e^{i\omega_{e}t}\,\big]}
\nonumber\\&&
\times\Upsilon_rc_{rk}e^{-i\omega_{rk}t}+\rm{H.c.}\big\},
\label{HT(t)-E}\\
H_{\rm ep}(t)\!&\!=\!&\!\left[ {\sin \theta \,\,\sigma_z^{(e)}+ \cos
\theta\,\,(\sigma_{+}^{(e)}e^{i\Omega_0t}+\sigma_{-}^{(e)}e^{-i\Omega_0t})}
\right]\nonumber\\&&\times\sum\limits_q {\lambda _q \left(
{b_q^{\dagger}e^{i\omega_qt} + b_qe^{-i\omega_qt}}
\right)},\label{Hep(t)-E}
\end{eqnarray}
where $\omega_{g,e}=\mp\Omega_0/2$. Here
$\sigma_{-}^{(e)}=a_ga_e^{\dagger}$ and
$\sigma_{+}^{(e)}=(\sigma_-^{(e)})^{\dagger}$ are the lowering and
raising operators in the eigenstate basis.

Now, one can evaluate Eqs.~(\ref{A-ME-O})--(\ref{Electron-Phonon})
using Eqs.~(\ref{HT(t)-E}) and (\ref{Hep(t)-E}). After converting
the result to the Schr\"{o}dinger picture, one obtains the master
equation in the eigenstate basis which holds for any arbitrary
interdot coupling $\Omega$ (see Appendix B):
\begin{eqnarray}
\dot{\rho}_d(t)\!&\!=\!&\!-i[\,H_{\rm
DQD},~\rho_d(t)\,]+\frac{\Gamma_L}{2}\alpha^2\mathcal{D}\,[a_g^{\dagger}]\,\rho_d
\nonumber\\
\!&\!\!&\!+\frac{\Gamma_L}{2}\beta^2\mathcal{D}\,[a_e^{\dagger}]\,\rho_d
+\frac{\Gamma_R}{2}\beta^2\mathcal{D}\,[a_g\Upsilon_r^{\dagger}]\rho_d
\nonumber\\
&&+\frac{\Gamma_R}{2}\alpha^2\mathcal{D}[a_e\Upsilon_r^{\dagger}]\rho_d
+\frac{\lambda_1}{2}\mathcal{D}[\sigma_-^{(e)}]\rho_d+\frac{\lambda_2}{2}[\sigma_+^{(e)}]\rho_d,
\nonumber\\\label{ME-E}
\end{eqnarray}
where
\begin{eqnarray}
\alpha\equiv\cos{\frac{\theta}{2}}=\sqrt{\frac{\Omega_0+\varepsilon}{2\Omega_0}},~~~
\beta\equiv\sin{\frac{\theta}{2}}=\sqrt{\frac{\Omega_0-\varepsilon}{2\Omega_0}},\label{alpha}
\end{eqnarray}
and
\begin{eqnarray}
\lambda_1&=&\gamma_0\cos^2{\theta}\,[\,n(\Omega_0)+1\,],\nonumber\\
\lambda_2&=&\gamma_0\cos^2{\theta}\,n(\Omega_0).\,\label{dissipation
rate-E}
\end{eqnarray}
with $\gamma_0=2\pi{J}(\Omega_0)$. Using Eq.~(\ref{ME-E}), the
$n$-resolved equation of motion for each density matrix element can
be written as
\begin{eqnarray}
\dot{\rho}_{00}^{(n)}(t)\!&\!=\!&\!-\Gamma_L\rho_{00}^{(n)}
+\Gamma_R\beta^2\rho_{gg}^{(n-1)}+\Gamma_R\alpha^2\rho_{ee}^{(n-1)},
\nonumber\\
\dot{\rho}_{gg}^{(n)}(t)\!&\!=\!&\!\Gamma_L\alpha^2\rho_{00}^{(n)}-\Gamma_R\beta^2\rho_{gg}^{(n)}
+\lambda_1\rho_{ee}^{(n)}-\lambda_2\rho_{gg}^{(n)},
\nonumber\\
\dot{\rho}_{ee}^{(n)}(t)\!&\!=\!&\!\Gamma_L\beta^2\rho_{00}^{(n)}-\Gamma_R\alpha^2\rho_{ee}^{(n)}
-\lambda_1\rho_{ee}^{(n)}+\lambda_2\rho_{gg}^{(n)}.\nonumber\\
&&\label{EOM-elements-E}
\end{eqnarray}
It follows from Eq.~(\ref{EOM-elements-E}) that the effective
tunneling rate from the left lead to the ground (excited) state
$|g\rangle$ ($|e\rangle$) is $\Gamma_L\alpha^2$ ($\Gamma_L\beta^2$),
while the effective tunneling rate from the ground (excited) state
to the right lead is $\Gamma_R\beta^2$ ($\Gamma_R\alpha^2$) [see
Fig.~1(b)]. We emphasize that these results derived in the
eigenstate basis are valid for any arbitrary interdot coupling.
\begin{figure}[tbp]
\includegraphics[width=2.8in,
bbllx=61,bblly=184,bburx=461,bbury=743]{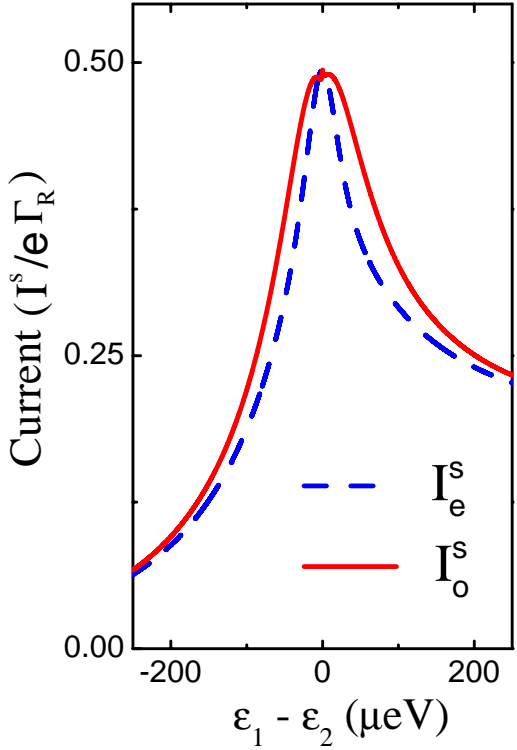} \caption{(Color
online)~Stationary current $I^s$ through the DQD as a function of
the energy difference $\varepsilon_1-\varepsilon_2$ calculated using
the occupation-state basis ($I^s=I_{\rm o}^s$) and the eigenstate
basis ($I^s=I_{\rm e}^s$) for a $large$ interdot coupling
$\Omega=32~\mu$eV. We have taken $\Gamma_L=100~\mu$eV,
$\Gamma_R=2.5~\mu$eV, $\gamma_0=0.6~\mu$eV, and $T=2~$K.
\label{fig2}}
\end{figure}

\section{current through the double quantum dot}

To compare the master equations in Eqs.~(\ref{EOM-elements-O}) and
~(\ref{EOM-elements-E}) derived, respectively, in the
occupation-state basis and the eigenstate basis, we first apply them
to study the tunneling current through the DQD. In the next section,
the associated shot noise will also be studied. The current $I(t)$
through the DQD at time $t$ is given by
\begin{eqnarray}
I(t)=e\frac{dN(t)}{dt}=e\sum_{n,i}n\dot{\rho}_{ii}^{(n)}(t),
\end{eqnarray}
where $N(t)$ is the number of electrons that have tunneled into the
right lead. Here, $i$ is summed over all basis states of the basis used.

Denoting results based on the occupation-state basis and the
eigenstate basis by ``o'' and ``e'', values of the current $I_{\rm
o}(t)$ and $I_{\rm e}(t)$ calculated using
Eqs.~(\ref{EOM-elements-O}) and (\ref{EOM-elements-E}) are
\begin{subequations}
\begin{align}
I_{\rm o}(t)\!&=\!e\,\Gamma_R\,\rho_{22}(t),\\
I_{\rm e}(t)\!&=\!e\,[\,\Gamma_R \beta^2 \rho_{gg}(t)  + \Gamma _R
\alpha ^2 \rho _{ee}(t)\,],
\end{align}
\end{subequations}
At steady-state with $\dot{\rho}_{ii}(t)=0$, calculated values
$I_{\rm o}^s$ and $I_{\rm e}^s$ of the stationary current are
\begin{subequations}
\begin{align}
I_{\rm o}^s\!&\!=\!\frac{e\Gamma_L \Gamma _R}{\Lambda}
\times\big\{\,\gamma _2 \;[\,4\varepsilon^2 +(\gamma _1 + \gamma_2 +
\Gamma_R)^2\,]
\nonumber\\
& + \;4\Omega^2(\gamma_1+\gamma_2+ \Gamma_R)\big\},
\label{stationary-current-a}\\
I_{\rm e}^s\!&\!=\!\frac{e{\Gamma _L \Gamma _R \left[ {\beta ^2
\lambda_1 + \alpha ^2 \left( {\lambda_2  + \beta ^2 \Gamma_R }
\right)} \right]}}{\Xi}, \label{stationary-current-b}
\end{align}
\end{subequations}
where
\begin{subequations}
\begin{align}
\Lambda \!&\!= 4\Omega^2 \Gamma _R%
\left( {2\Gamma _L  + \Gamma _R } \right)+ \left( {\gamma _2  +
\Gamma _L } \right)\Gamma _R \left( {4\varepsilon ^2  + \Gamma _R^2} \right)
\nonumber\\
& + \left( {\gamma _1  + \gamma _2 } \right)\bigg[4\Gamma_L(
\varepsilon ^2 + 2\Omega ^2) + 4\Omega^2 \Gamma_R
\nonumber\\
& + \Gamma_L (\gamma_1  + \gamma_2)^2+\,\Gamma _R (\gamma_2  +
3\Gamma _L)(\gamma_1 + \gamma_2) \nonumber\\&
+ \Gamma_R^2\,(2\gamma_2 + 3\Gamma_L ) \bigg],
\label{Lambda}\\
\Xi\!&\!=\lambda_2 \left(\Gamma _L  + \alpha ^2 \Gamma_R\right) +
\lambda_1 \left( \Gamma_L + \beta ^2 \Gamma _R\right)
\nonumber\\
&+ \Gamma_R \left[\left(\alpha ^4  + \beta ^4 \right)\Gamma_L +
\alpha ^2 \beta ^2 \Gamma_R  \right].\label{Xi}
\end{align}
\end{subequations}

Figure \ref{fig2} shows the calculated values $I_{\rm o}^s$ and
$I_{\rm e}^s$ of the stationary current through the DQD. We choose a
typical interdot coupling $\Omega=32~\mu$eV, which is experimentally
accessible.\cite{Gustavsson07} For both $I_{\rm o}^s$ and $I_{\rm
e}^s$ at the resonant tunneling point characterized by
$\varepsilon_1-\varepsilon_2=0$, the current reaches its maximum.
Moreover, it can be seen that the current is asymmetric around the
maximum point. This asymmetry was also observed in a recent
experiment by Barthold {\it et al}..\cite{Barthold07} It is due to
dissipations induced by the phonon bath, as we will now demonstrate.
In the absence of electron-phonon coupling, we have
$\gamma_1=\gamma_2=0$ and $\lambda_1=\lambda_2=0$, so that
Eqs.~(\ref{stationary-current-a}) and (\ref{stationary-current-b})
reduces to
\addtocounter{equation}{1}
\begin{align}
I_{\rm o}^s\!&=\!\frac{{4\,e\,\Omega ^2 \Gamma _L \Gamma _R
}}{{4\Omega ^2 \left( {2\Gamma _L  + \Gamma _R } \right) +
4\varepsilon ^2 \Gamma _L  + \Gamma _L \Gamma _R^2
}},\label{I-stationary-a}
\tag{\theequation a}\\
I_{\rm e}^s\!&=\!\frac{{4\,e\,\Omega ^2\Gamma _L \Gamma _R
}}{{4\Omega ^2 \left( {2\Gamma _L  + \Gamma _R } \right) +
4\varepsilon ^2 \Gamma _L }},\tag{\theequation b}
\label{I-stationary-b}
\end{align}
\begin{figure}[tbp]
\includegraphics[width=2.80in,
bbllx=63,bblly=178,bburx=462,bbury=741]{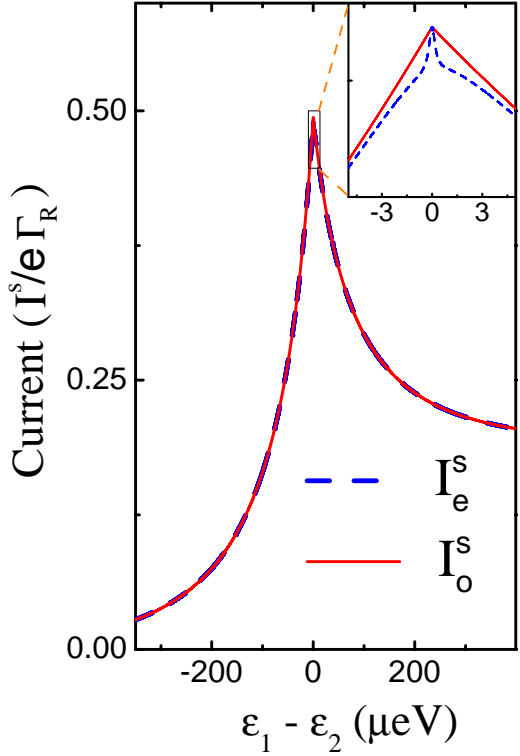} \caption{(Color
online)~~Stationary current $I^s$ as a function of
$\varepsilon_1-\varepsilon_2$ calculated using the occupation-state
basis ($I^s=I_{\rm o}^s$) and the eigenstate basis ($I^s=I_{\rm
e}^s$) for a $small$ interdot coupling $\Omega=1~\mu$eV. Inset: The
enlarged diagram of the stationary current $I^s$ in the region with
$|\varepsilon_2-\varepsilon_1|$ comparable to $\Omega$.\label{fig3}
}
\end{figure}
where Eq.~(\ref{I-stationary-a}) agrees with the result from
previous studies,\cite{Gurvitz02,Kiesslich07} in which the
occupation-state basis was also used.  From
Eqs.~(\ref{I-stationary-a}) and (\ref{I-stationary-b}), it is clear
that the current as predicted using either the occupation-state
basis or the eigenstate basis is symmetric about the current peak at
$\varepsilon_1=\varepsilon_2$. This reveals that this asymmetry of
the current is due to the coupling of the DQD to the phonon bath.

As shown in Sec.~II, the master equation derived in the
occupation-state basis is only valid in the limit of a weak interdot
coupling, i.e., $\Omega\ll|\varepsilon_2-\varepsilon_1|$, while that
in the eigenstate basis is valid for any arbitrary interdot
coupling. Figure~{\ref{fig3}} plots the values of the stationary
current calculated in the two bases for a small interdot coupling
($\Omega=1~\mu$eV). As expected, when the interdot coupling $\Omega$
is much smaller than the energy difference
$|\varepsilon_2-\varepsilon_1|$ of the two dots, the
stationary-state current calculated in the occupation-state basis
agrees very well with that in the eigenstate basis, as is evident
from Fig.~\ref{fig3}. However, when the interdot coupling is
comparable to the energy difference, the stationary currents in the
occupation-state basis deviates drastically from the stationary
current in the eigenstate basis (see Fig.~\ref{fig2}). This
deviation can also be revealed in Fig.~\ref{fig3} in the narrow
region with $|\varepsilon_2-\varepsilon_1|\sim\Omega$~($=1\mu eV$)
(see the inset of Fig.~\ref{fig3}). These clearly show the
inaccuracy of the current and hence the master equation in the
occupation-state basis at large $\Omega$. In this case, one must use
the master equation derived in the eigenstate basis.

\section{shot noise}

To calculate the shot noise in the tunneling current through the
DQD, it is particularly useful to define a generating function for
an electron counting variable $s$ (see
Refs.~\onlinecite{Sanchez07,Ouyang08} and \onlinecite{DongPRL}):
\begin{equation}
G(t,s)= \sum_{n}s^{n}\rho^{(n)}(t).\label{Generating Function}
\end{equation}
This generation function obeys the equation of motion
\begin{eqnarray}
\dot{G}(t,s)=M(s)G(t,s),\label{GF}
\end{eqnarray}
where $M(s)$ is a transition matrix that can be calculated using the
master equation [Eq.~(\ref{ME-O}) or (\ref{ME-E})]. Statistics on
the number of transported electrons $n$ can be determined from the
derivatives of the generating function:
\begin{equation}
\frac{\partial^{p}{\rm{tr}}G(t,1)}{\partial
s^p}=\big\langle\prod_{i=1}^{p}(n-i+1)\big\rangle.
\end{equation}
In particular, the mean of $n$ is
\begin{equation}
\langle n\rangle=\frac{\partial {\rm{tr}}G(t,1)}{\partial s},
\end{equation}
and the variance reads
\begin{eqnarray}
\sigma_{n}^2=\langle n^2\rangle-\langle n\rangle^2 =\frac{\partial^2
{\rm{tr}} G(t,1)}{\partial s^2}+\langle n\rangle-\langle n\rangle^2.
\end{eqnarray}
Applying the Laplace transform to the equation of motion,
Eq.~(\ref{GF}), of the generating function, one has
\begin{equation}
\tilde{G}(z,s)=(z-M)^{-1}G(0,s).\label{LP}
\end{equation}
Because of the incoherent long-time stability of the considered
system, the real parts of all the non-zero poles of $\tilde{G}(z,s)$
are negative. Therefore, the long-time behavior is determined by the
pole $z_0$ closest to zero, i.e., $G(t,s)\!\sim\!g(s)e^{z_0t}$. By
the Taylor expansion of the pole
\begin{equation}
z_0=\sum_{m>0}c_{m}(s-1)^m,\label{Taylor}
\end{equation}
one obtains
\begin{eqnarray}
&&\langle n\rangle=\frac{\partial{g(1)}}{\partial s}+c_{1}t,
\nonumber\\
&&\sigma_{n}^2= \frac{\partial^2 g(1)}{\partial{s^2}}
-\bigg(\frac{\partial g(1)}{\partial s}\bigg)^2
+(c_{1}+2c_{2})t.\nonumber\\
&&~~~~~~~~~~~~~~~~~~~~~~~~~~~~~~~~~~~~~\label{Fano factor}
\end{eqnarray}
In particular, the Fano factor of the shot noise is given by
\begin{equation}
F=1+2\;\frac{c\,_{2}}{c\,_{1}},\label{F}
\end{equation}
where $F>1$~$(F<1)$ indicates super(sub)-Poissonian noise, compared
to $F=1$ for classical Poissonian noise.

We first consider results based on the occupation-state basis. To
calculate the Fano factor, we can show, using
Eqs.~(\ref{EOM-elements-O}) and (\ref{LP}), that the pole $z_0$
follows
\begin{eqnarray}
&&a_1 \left( {s  - 1} \right) + a_2 z_0+ a_3 \left( {s  - 1}
\right)z_0 \nonumber\\
&& +\; a_4 z_0^2  + a_5 z_0^2 \left( {s  - 1} \right) + a_6 z_0^3 +
a_7 z_0^4 + z_0^5=0, \nonumber\\ &&\label{pole-O}
\end{eqnarray}
with
\begin{eqnarray}
a_1\!&\!=\!&\! - \frac{1}{4}\Gamma_L \Gamma_R \big\{4\varepsilon^2 \gamma_2 %
+ (\gamma_1  + \gamma_2 + \Gamma _R)
\nonumber\\
&&\times[4\Omega^2 + \gamma_2(\gamma_1 + \gamma_2 +
\Gamma_R)]\big\}, \nonumber\\
a_2\!&\!=\!&\!
\frac{1}{4}\bigg\{\Gamma_L(\gamma_1+\gamma_2)[4(\varepsilon^2+2\Omega^2)+(\gamma_1+\gamma_2)^2]
\nonumber\\
&&+\big\{(\gamma_1+\gamma_2)[4\Omega^2+\gamma_2(\gamma_1+\gamma_2)]
\nonumber\\&&
+\Gamma_L[8\Omega^2+3(\gamma_1+\gamma_2)^2]+4\varepsilon^2(\gamma_2+\Gamma_L)\big\}\Gamma_R
\nonumber\\&&
+[4\Omega^2+(\gamma_1+\gamma_2)(2\gamma_2+3\Gamma_L)]\Gamma_R^2+(\gamma_2+\Gamma_L)\Gamma_R^3\bigg\},
\nonumber\\
a_3\!&\!=\!&\!-\Gamma_L\Gamma_R[2\Omega^2+\gamma_2(\gamma_1+\gamma_2+\Gamma_R)],\nonumber\\
a_4\!&\!=\!&\!\frac{1}{4}\bigg\{\gamma_1^3+\gamma_2^3+4(\varepsilon^2+4\Omega^2)(\Gamma_L+\Gamma_R)
+5\Gamma_L\Gamma_R^2\nonumber\\
&&
+\Gamma_R^3+\gamma_1^2(3\gamma_2+5\Gamma_L+3\Gamma_R)+\gamma_2^2(5\Gamma_L+7\Gamma_R)
\nonumber\\&&
+\gamma_2(4\varepsilon^2+8\Omega^2+10\Gamma_L\Gamma_R+7\Gamma_R^2)
\nonumber\\
&&+\gamma_1\big[4\varepsilon^2+8\Omega^2+3\gamma_2^2+10\Gamma_L\Gamma_R+3\Gamma_R^2
\nonumber\\
&&+10\gamma_2(\Gamma_L+\Gamma_R)\big] \bigg\}.
\end{eqnarray}
The expressions for $a_5$, $a_6$ and $a_7$ are not involved in further
calculations and are not reported here.
\begin{figure}[tbp]
\includegraphics[width=2.8in,
bbllx=63,bblly=170,bburx=460,bbury=750]{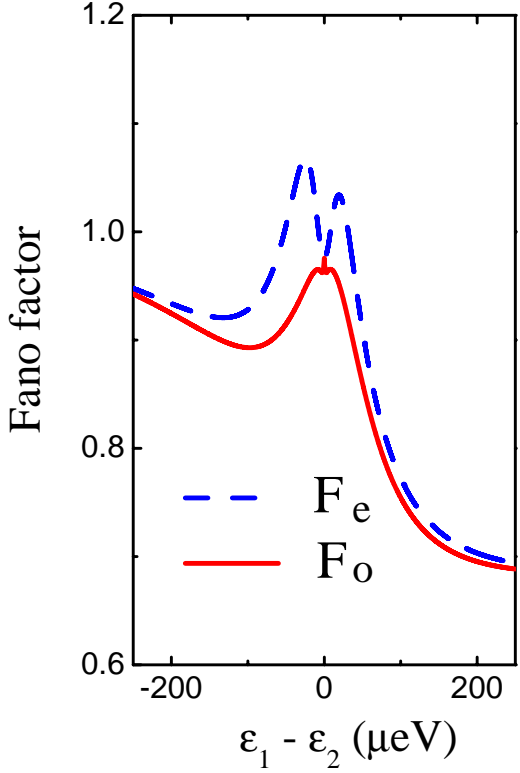} \caption{(Color
online)~Fano factor $F$ as a function of
$\varepsilon_1-\varepsilon_2$ calculated using the occupation-state
basis ($F=F_{\rm o}$) and the eigenstate basis ($F=F_{\rm e}$) for a
$large$ interdot coupling $\Omega=32~\mu$eV. \label{fig4}}
\end{figure}
Using also Eqs.~(\ref{Taylor}) and (\ref{F}), the Fano factor
$F_{\rm o}$ in the occupation-state basis is found to be
\begin{equation}
F_{\rm o}=1+2\times\frac{a_1\,a_4\,-\,a_2\,a_3}{a_2^2}.\label{FO}
\end{equation}
Without any phonon dissipation effect, i.e., $\gamma_1=\gamma_2=0$, the
Fano factor becomes
\begin{equation}
F_{\rm o}\!=\!1 - \frac{{8\Omega ^2 \Gamma _L \left[ {4\varepsilon
^2 \left( {\Gamma _R  - \Gamma _L } \right) + 3\Gamma _L \Gamma _R^2
+ \Gamma _R^3  + 8\Omega ^2 \Gamma _R } \right]}}{{\left[ {\Gamma _L
\Gamma _R^2  + 4\Gamma _L \varepsilon ^2  + 4\Omega ^2 \left(
{\Gamma _R  + 2\Gamma _L } \right)} \right]^2 }},\label{FO-no
phonon}
\end{equation}
which is identical to the previous results\cite{Gurvitz02,Brandes05}
obtained in the occupation-state basis. For a super-Poissonian
noise, one has $F_{\rm o}>1$. From Eq.~(\ref{FO-no phonon}), it
follows that
\begin{equation}
4\varepsilon^2(\Gamma_L-\Gamma_R)>3\Gamma _L \Gamma _R^2 + \Gamma
_R^3  + 8\Omega ^2 \Gamma _R.\label{Condition}
\end{equation}

Alternatively, using the eigenstate basis, one can obtain from
Eqs.~(\ref{EOM-elements-E}) and (\ref{LP}) the following equation
for the pole $z_0$:
\begin{equation}
b_1(s-1)+b_2z_0+b_3z_0+b_4z_0+z^3=0,\label{pole-E}
\end{equation}
where
\begin{eqnarray}
b_1\!&\!=\!&\!-\Gamma_L\Gamma_R\,[\alpha^2(\lambda_2+\beta^2\Gamma_R)+\beta^2\lambda_1\,],
\nonumber\\
b_2\!&\!=\!&\!\lambda_1(\Gamma_L+\Gamma_R\beta^2)+\lambda_2(\Gamma_L+\alpha^2\Gamma_R)
\nonumber\\
&&+\,\Gamma_R[\,\Gamma_L(\alpha^2+\beta^2-2\alpha^2\beta^2)+\alpha^2\beta^2\Gamma_R\,],
\nonumber\\
b_3\!&\!=\!&\!-2\alpha^2\beta^2\Gamma_L\Gamma_R,\nonumber\\
b_4\!&\!=\!&\!\lambda_1+\lambda_2+\Gamma_L+\Gamma_R.
\end{eqnarray}
In contrast to Eq.~(\ref{pole-O}), only four coefficients $b_i$
($i=1$ to $4$) appear in Eq.~(\ref{pole-E}). From an equation for
$F_{\rm e}$ analogous to Eq.~(\ref{FO}) for $F_{\rm o}$, we get
\begin{eqnarray}
F_e&=&1 + \frac{2\Gamma_L \Gamma_R}{\Xi^2} \times \big\{2\alpha ^2
\beta ^2 \,\Xi \,- ( \lambda _1 + \lambda_2  + \Gamma_L + \Gamma_R)
\nonumber\\
&&\times [\,\beta ^2  \lambda_1 + \alpha^2  \lambda_2  + \alpha^2
\beta^2 \Gamma_R \,]\,\big\},\label{FE}
\end{eqnarray}
where $\Xi$ is given in Eq.~(\ref{Xi}). Without phonon
dissipation, i.e., $ \lambda_1= \lambda_2=0$, one obtains after substituting
Eq.~(\ref{alpha}) into Eq.~(\ref{FE}),
\begin{equation}
F_{\rm e}  = 1 - \frac{{8\Omega ^2 \Gamma _L }\times \left\{
{4\varepsilon ^2 \left( {\Gamma _R  - \Gamma _L } \right) + 8\Omega
^2 \Gamma _R } \right\}}{{\left[\, {4\varepsilon ^2 \Gamma _L  +
4\Omega ^2 \left( {2\Gamma _L  + \Gamma _R } \right)} \,\right]^2
}}.\label{FE-no phonon}
\end{equation}
\begin{figure}[tbp]
\includegraphics[width=2.8in,
bbllx=63,bblly=180,bburx=461,bbury=749]{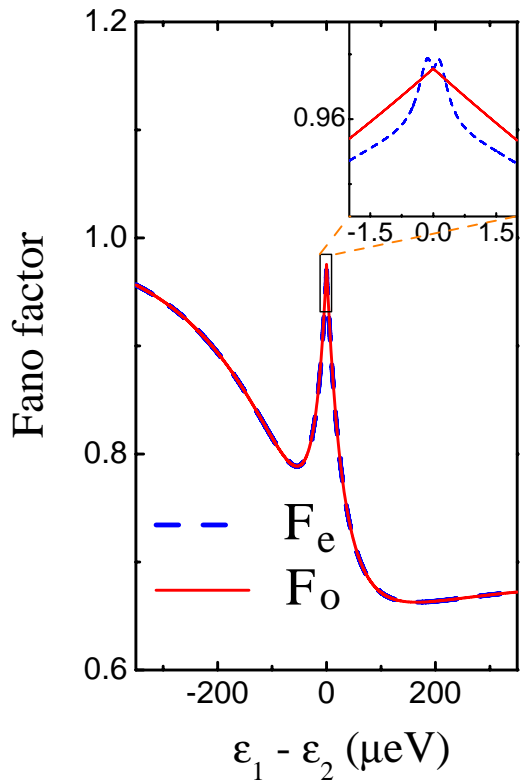} \caption{(Color
online)~Fano factor $F$ as a function of
$\varepsilon_1-\varepsilon_2$ calculated using the occupation-state
basis ($F=F_{\rm o}$) and the eigenstate basis ($F=F_{\rm e}$) for a
$small$ interdot coupling $\Omega=1~\mu$eV. Inset: The enlarged
diagram of the Fano factor $F$ in the region with
$|\varepsilon_1-\varepsilon_2|$ comparable to $\Omega$.
\label{fig5}}
\end{figure}
Figure \ref{fig4} presents both of the calculated Fano factors
$F_{\rm o}$ and $F_{\rm e}$ of the shot noise based on the
occupation-state basis and eigenstate basis, respectively. At the
resonant tunneling point, i.e., $\varepsilon_1=\varepsilon_2$, both
approaches predict that the shot noise is sub-Poissonian. For
$F_{\rm o}$, in the absence of phonon-induced dissipation, a
super-Poissonian noise can be obtained with the condition
Eq.~(\ref{Condition}). Due to the effects of dissipation, $F_{\rm
o}$ has only sub-Poissonian noise for the whole parameter range
investigated here. In contrast, $F_{\rm e}$ has much richer
behaviors of super-Poissonian, sub-Poissonian, and Poissonian noise
correlations, depending on the
energy difference $\varepsilon_1-\varepsilon_2$. 
Moreover, $F_{\rm e}$, but not $F_{\rm o}$, exhibits a double-peak
structure and an asymmetry around the dip at
$\varepsilon_1=\varepsilon_2$. These features were also observed in
a recent experiment (see Ref.~\onlinecite{Barthold07}).

The double peak in the Fano factor predicted using the eigenstate
basis can be intuitively understood as follows. The electrons can
tunnel from the DQD to the right lead via two channels, namely, the
ground-state channel and the excited-state channel. At the resonant
tunneling point ($\varepsilon_1=\varepsilon_2$), the tunnel rate
through the ground-state channel is the same as that through the
excited-state channel. This results in a sub-Poissonian shot
noise.\cite{Sanchez07} When $\varepsilon_1<\varepsilon_2$, one has
$\Gamma_g<\Gamma_e$, and the electron transport through the
ground-state channel blocks that through the excited-state channel.
This dynamical channel blockade leads to a super-Possionian shot
noise.\cite{Belzig05} However, when
$\varepsilon_2-\varepsilon_1\gg\Omega$, $\Gamma_g$ becomes zero and
the electron can only tunnel through the DQD via the excited-state
channel. This single-channel tunneling gives rise to a
sub-Poissonian shot noise,\cite{Chen92} as shown in Fig.~\ref{fig4}.
Similarly, when $\varepsilon_1>\varepsilon_2$, one has
$\Gamma_g>\Gamma_e$ and the tunneling through the excited-state
channel blocks that through the ground-state channel. The noise is
super-Poissonian for a small energy difference
$\varepsilon_1-\varepsilon_2$, due to the dynamical channel
blockade. When $\varepsilon_1-\varepsilon_2\gg\Omega$, $\Gamma_e$
becomes zero and the electron can only tunnel through the DQD via
the ground-state channel. This single-channel tunneling also gives
rise to a sub-Poissonian noise.

The asymmetry of the shot noise is caused by the relaxation process
induced by the electron-phonon interaction. When
$\varepsilon_1<\varepsilon_2$, we have $\Gamma_g<\Gamma_e$ and the
relaxation process from the excited state to the ground state {\it
enhances} the dynamical channel blockade. However, when
$\varepsilon_1>\varepsilon_2$, one has $\Gamma_g>\Gamma_e$ and the
relaxation process from the excited state to the ground state {\it
suppresses} the dynamical channel blockade. The asymmetry of the
Fano factor hence follows [see Fig.~\ref{fig4}].

We have shown using Fig.~\ref{fig4} that for a large interdot
coupling, e.g., $\Omega=32~\mu$eV, the Fano factor in the
occupation-state basis deviates drastically from the Fano factor in
the eigenstate basis. This verifies that the small interdot coupling
approximation in deriving the master equation in the
occupation-state basis is invalid. Instead, the master equation in
the eigenstate basis should be used. As a further consistency check,
Fig. \ref{fig5} shows the Fano factor of the shot noise for a small
interdot coupling strength ($\Omega=1~\mu$eV). As expected, the
calculated Fano factors using both basis agrees with each other
except for the small region with
$|\varepsilon_2-\varepsilon_1|\sim\Omega$; in this small region, the
results in the two cases are different because the condition
$|\varepsilon_2-\varepsilon_1|\gg\Omega$ is not satisfied (see the
inset of Fig.~\ref{fig5}).

\section{Correction terms for the Occupation-State Master equation}

In this section, we derive a controlled series expansion for the
scattering term in the quantum master equation with respect to the
interdot coupling strength. This provides a concise quantitative
description of the approximation used in the occupation-state
approach. In general, it can also allow one to derive correction
terms, either to improve the results based on the occupation-state
approach or to estimate the resulting error.

The master equations in both approaches are derived from
Eq.~(\ref{ME-1}) in the interaction picture. 
%
%
To study the difference between the two approaches, we first
transform Eq.~(\ref{ME-1}) back to the Schr\"{o}dinger picture and
get
\newcommand{\inttau}{\int_{0}^\infty d\tau}
\begin{eqnarray}
\dot{\rho}_{ d}(t) \!&\!=\!&\!
-i{\rm Tr}_E[H_0, \rho_d(t)\rho_E(0)] \nonumber \\
&& - {\rm Tr}_E\int\limits^\infty_0d\tau [H_{\rm int},
e^{-iH_0\tau}[\,H_{\rm int}, \rho_d(t)\rho_E(0)]
e^{iH_0\tau}],\nonumber\\&&\label{ME-3}
\end{eqnarray}
where we have put $\tau=t-t'$ and assumed $t\gg 0$. It can further
be written in the more compact form \cite{Weiss}
\begin{eqnarray}
\dot{\rho}_{ d}(t)\!&=&\! {\rm Tr}_E[\Li_0 \rho_d(t)\rho_E]
\nonumber \\&&
+ {\rm Tr}_E\int\limits^\infty_0d\tau \Li_\Int
e^{\Li_0\tau} \Li_\Int \rho_d(t)\rho_E,  \label{ME-4}
\end{eqnarray}
where $\Li_0$ and $\Li_\Int$ are the Liouville operators for the
Hamiltonians $H_0$ and $H_\Int$, respectively. The Liouville
operator $\Li_0$, for instances, is defined by $\Li_0 A = -i [ H_0 ,
A]$ for any operator $A$. We have also used $e^{\Li_0\tau}A =
e^{-iH_0\tau}Ae^{iH_0\tau}$ which follows directly from the
Baker-Hausdorff lemma.\cite{Sakurai}

In the eigenstate basis, $e^{\Li_0\tau}$ in Eq.~(\ref{ME-4}) is
treated exactly. However, in the occupation-state basis, it can only
be approximated. To illustrate this approximation and study the
associated correction terms, we note that $\Li_0 = \Li_1 +
\Li_\Omega$ because of Eq.~(\ref{H-0}) and derive the Dyson series
\begin{equation}
  \label{Li0}
  e^{\Li_0\tau} =   e^{\Li_1\tau} + \int_0^{\tau}d\tau'
  e^{\Li_1(\tau-\tau')} \Li_\Omega e^{\Li_1\tau'} +\cdots,
\end{equation}
where $\Li_1$ and $\Li_\Omega$ denote the Liouville operators for
$H_1$ and $H_\Omega$, respectively. Equation (\ref{ME-4}) then
becomes
\begin{eqnarray}
&&\dot{\rho}_{ d}(t)={\rm Tr}_E[\Li_0 \rho_d(t)\rho_E]
\nonumber\\
&&+ {\rm Tr}_E\int_{0}^\infty d\tau \Li_\Int e^{\Li_1\tau} \Li_\Int
\rho_d(t)\rho_E
 \nonumber \\
&& + {\rm Tr}_E\int\limits_{0}^\infty d\tau\!\int\limits_{0}^\tau
\!d\tau' \Li_\Int e^{\Li_1(\tau-\tau')} \Li_{\Omega} e^{\Li_1\tau'}
\Li_\Int
\rho_d(t)\rho_E   \nonumber \\
&& +\cdots. \label{ME-5}
\end{eqnarray}
Taking only the first two terms, we arrive at the approximate master
equation used in the occupational-state basis, which is identical to
Eq.~(\ref{ME-4}) with $e^{\Li_0\tau}$ approximated by
$e^{\Li_1\tau}$. The third term in Eq.~(\ref{ME-5}) is then the
leading correction term for the master equation in the
occupational-state basis. It consists of terms of order $\mathcal
O(\Omega\Omega_{lk}^2)$ or $\mathcal O(\Omega\lambda_q^2)$.
Expressions for higher order correction terms in the
occupational-state approach can similarly be calculated.

For our DQD problem, the correction terms can also be obtained by a
direct comparison with the eigenstate basis result. Without the lose
of generality, we assume $\varepsilon_2-\varepsilon_1>0$ in the
following discussion. In the small interdot-coupling limit with
$\Omega\ll|\varepsilon_2-\varepsilon_1|$, $\alpha$ and $\beta$
defined in Eq.~(\ref{alpha}) reduce approximately to
\begin{eqnarray}
\alpha\approx1, ~~\beta\approx\eta,
\end{eqnarray}
where $\eta=\Omega/\varepsilon\ll1$. The transformation between the
two bases given in Eq. (\ref{basis-transform}) can be
approximated by
\begin{eqnarray}
|g\rangle\!&\!\approx\!&\!|1\rangle-\eta\,|2\rangle,\nonumber\\
|e\rangle\!&\!\approx\!&\!\eta\,|1\rangle+|2\rangle.\label{Eigenstate-Approx}
\end{eqnarray}
Substituting Eq.~(\ref{Eigenstate-Approx}) into the master equation
in the eigenstate basis [Eq.~(\ref{ME-E})], and keeping terms only up to first order in
$\eta$, one has
\begin{eqnarray}
&&\dot{\rho}_d(t)\approx-i\big[\,%
H_{\rm DQD} ,~\rho_d(t)\,\big]
+\frac{\Gamma_L}{2}\mathcal{D}[a_1^{\dagger}]\rho_d
\nonumber\\
&&+\frac{\Gamma_R}{2}\mathcal{D}[a_2\Upsilon_r^{\dagger}]\rho_d
+\frac{\gamma_1}{2}\mathcal{D}[a_2^{\dagger}a_1]\rho+\frac{\gamma_2}{2}\mathcal{D}[a_1^{\dagger}a_2]\rho_d
\nonumber\\
&&-\Gamma_L\eta[a_2^{\dagger}\rho_d a_1 + a_1^{\dagger}\rho_d a_2]
\nonumber\\&&
+\frac{\Gamma_R}{2}\eta\big[2a_1\Upsilon_r^{\dagger}\rho_d\Upsilon_r
a_2^{\dagger}+2a_2\Upsilon_r^{\dagger}\rho_d\Upsilon_r
a_1^{\dagger}-\sigma_x\rho_d-\rho_d\sigma_x\big]
\nonumber\\
&&-\gamma_1\eta\big[a_1^{\dagger}a_2\rho_d\sigma_z+\sigma_z\rho_d
a_2^{\dagger}a_1\big]
-\frac{\gamma_1-\gamma_2}{2}\eta[\sigma_x\rho_d+\rho_d\sigma_x]
\nonumber\\&&
-\gamma_2\eta\big[a_2^{\dagger}a_1\rho_d\sigma_z+\sigma_z\rho_d
a_1^{\dagger}a_2\big].\label{ME-E-approx}
\end{eqnarray}
Indeed, when $\eta\!=\!0$, this equation reduces to the approximate
master equation in the occupation-state basis [Eq.~(\ref{ME-O})].
The terms proportional to $\eta$ are the leading correction terms of
the order $\mathcal{O}(\Omega\Omega_{lk}^2)$ or
$\mathcal{O}(\Omega\lambda_q^2)$ as expected.

\section{Conclusion}

In summary, we have derived two master equations in both the
occupation-state basis and the eigenstate basis to describe the
dynamics of the DQD. We show that the master equation in the
occupation-state basis is only valid for a small interdot coupling,
while the master equation in the eigenstate basis is valid for an
arbitrary interdot coupling. To demonstrate the difference between
these two master equations, we focus on the current and shot-noise
properties in electron tunneling through the DQD. When the interdot
coupling is much smaller than the energy difference between the two
dots, the current and shot noise in the occupation-state basis are
very close to those in the eigenstate basis. For a large interdot
coupling, however, the properties derived in the occupation-state
basis deviate drastically from those in the eigenstate basis. This
reveals that the master equation in the occupation-state basis is
not accurate for the case of a large interdot coupling and in this
case the master equation in the eigenstate basis should be used.
Also, we show that the shot-noise properties predicted using the
eigenstate basis can successfully reproduce the features found in a
recent experiment.\cite{Barthold07} Moreover, we have discussed the
relation between these two master equations and show explicitly that
the master equation in the occupation-state basis only includes low
order terms with respect to the interdot coupling, compared with the
master equation derived in the eigenstate basis.

\begin{acknowledgments}
This work is supported by the National Basic Research Program of
China Grant Nos. 2009CB929300 and 2006CB921205, the National Natural
Science Foundation of China Grant Nos. 10534060 and 10625416, and
the Research Grant Council of Hong Kong SAR project No. 500908.
\end{acknowledgments}

\appendix
\section{Derivation of master equation in occupation-state basis}

In this appendix, we give further details of the derivation of the master equation in the
occupation-state basis outlined in Sec. IIA.
We first evaluate $\mathcal{L}_{\rm T}\rho_d^I(t)$. Using the expression for $H_T(t)$ in
Eq.~(\ref{H-T(t)}),  the first term in
Eq.~(\ref{Tunneling}) becomes
\begin{eqnarray}
&&-\int_{0}^\infty d\tau\bigg\{\sum\limits_{lk} {\Omega_{lk}^2 a_1
a_1^{\dagger}\rho _d^I \left( t \right)
e^{i\left( {\omega _{lk} -\omega _1 } \right)\tau } \left\langle
{c_{lk}^{\dagger} c_{lk} } \right\rangle }
\nonumber\\
&&+\sum\limits_{lk}{\Omega _{lk}^2 a_1^{\dagger}  a_1 \rho _d^I
\left( t \right)
e^{ -i\left( {\omega _{lk}  - \omega _1 }\right)\tau }
\left\langle {c_{lk} c_{lk}^{\dagger}  } \right\rangle }  \nonumber\\
&&+ \sum\limits_{rk} {\Omega _{rk}^2 a_2 a_2^{\dagger} \Upsilon _r^
{\dagger} \Upsilon _r \rho _d^I \left( t \right)
e^{i\left( {\omega_{rk}  - \omega _2 } \right)\tau } \left\langle
{c_{rk}^{\dagger}
c_{rk} } \right\rangle } \nonumber \\
&&+ \sum\limits_{rk} {\Omega _{rk}^2 a_2^{\dagger}  a_2 \Upsilon _r
\Upsilon _r^{\dagger} \rho _d^I \left( t \right)
e^{- i\left( {\omega _{rk}  - \omega _2 } \right)\tau } \left\langle
{c_{rk} c_{rk}^{\dagger}  } \right\rangle }
\bigg\},\nonumber\\&&\label{A9}
\end{eqnarray}
where $\tau=t-t'$. When the electron density of states in an
electric lead is dense, each sum in Eq.~(\ref{A9}) can be replaced
by an integral. After some algebra, we obtain
\begin{eqnarray}
&&-\sum_{\alpha=l,r}{\rm
Tr}_{\alpha}\int\limits_0^{t}dt'\big[\,H_{\rm T}(t)H_{\rm
T}(t')\rho_d^I(t)\,\rho_{\rm leads}(0) \nonumber\\&&
=- \frac{{\Gamma _L }}{2}\left[ {a_1 a_1^{\dagger} \rho _d^I f_l
\left( {\omega _1 } \right) + a_1^{\dagger}  a_1  \rho _d^I \bar f_l
\left( {\omega _1 } \right)} \right] \nonumber\\&&
- \frac{{\Gamma _R }}{2}\left[ {a_2 a_2^{\dagger}  \Upsilon
_r^{\dagger} \Upsilon _r \rho _d^I f_r \left( {\omega _2 } \right) +
a_2^{\dagger}  a_2 \Upsilon _r \Upsilon _r^{\dagger}  \rho _d^I \bar
f_r \left( {\omega _2 } \right)} \right], \nonumber \\&& \label{A10}
\end{eqnarray}
where $\Gamma_{L,R}=2\pi\rho_{lr,rk}\Omega_{lk,rk}^2$ is the
electron tunneling rate through the left (right) barrier. Here
\begin{equation}
f_{\alpha}(\omega_i)=\frac{1}{1+e^{(\omega_i-\mu_{\alpha})/k_BT}},
\end{equation}
is the Fermi-Dirac distribution with $\mu_\alpha$ being the chemical
potential of lead $\alpha$ and $\bar
f_\alpha(\omega_i)=1-f_\alpha(\omega_i)$. Note that, in deriving
Eq.~(\ref{A10}), we have used the relations
\begin{equation}
\langle c_{\alpha k}^{\dagger}c_{\alpha
k}\rangle=f_{\alpha}(\omega_{\alpha k}),~~~ \langle c_{\alpha
k}c_{\alpha k}^{\dagger}\rangle=1-f_{\alpha}(\omega_{\alpha k}),
\end{equation}
and
\begin{equation}
\int_0^{\infty}d\tau \;e^{\pm i(\omega_{\alpha
k}-\omega_i)\tau}\approx\pi\delta(\omega_{\alpha k}-\omega_i).
\end{equation}
Similarly, the second term in Eq.~(\ref{Tunneling}) can be
calculated as
\begin{eqnarray}
&&\sum_{\alpha=l,r}{\rm
Tr}_{\alpha}\int\limits_0^{t}dt'\big[\,H_{\rm
T}(t)\rho_d^I(t)\,\rho_{\rm leads}(0)H_{\rm T}(t')\,\big]
\nonumber\\&& =\frac{{\Gamma _L }}{2}\left[ {a_1 \rho
_d^Ia_1^{\dagger} \bar f_l \left( {\omega _1 } \right)
+ a_1^{\dagger}  \rho _d^I a_1 f_l \left( {\omega _1 } \right)} \right]\nonumber \\
&&+ \frac{{\Gamma _R }}{2}\left[ {a_2 \Upsilon _r^{\dagger}  \rho
_d^I \Upsilon _r a_2^{\dagger}  \bar f_r \left( {\omega _2 } \right)
+ a_2^{\dagger}  \Upsilon _r \rho _d^I \Upsilon _r^{\dagger}
a_2 f_r \left( {\omega _2 } \right)} \right]. \nonumber\\
&& \label{A12}
\end{eqnarray}
Substituting
Eqs.~(\ref{A10}) and (\ref{A12}) into Eq.~(\ref{Tunneling}), one
obtains
\begin{eqnarray}
&&\mathcal{L}_{\rm T}\rho_d^I(t)= \frac{{\Gamma _L }}{2}\mathcal{D}[
a_1 ] \rho _d^I(t) \bar f_l ( \omega_1)+ \frac{{\Gamma _L }}{2}
\mathcal{D}[ a_1^{\dagger} ]\rho _d^I(t) f_l ( \omega
_1)\nonumber\\&&
 + \frac{{\Gamma _R }}{2}\mathcal{D}[ a_2 \Upsilon
_r^{\dagger}] \rho_d^I(t) \bar f_r \left( {\omega_2 } \right)
+ \frac{{\Gamma _R }}{2} \mathcal{D}\left[ {a_2^{\dagger}  \Upsilon
_r } \right]\rho _d^I(t) f_r \left( {\omega _2 } \right),
\nonumber\\&&\label{L-T}
\end{eqnarray}
where $\mathcal{D}$ (acting on any operator $A$) is defined by
\begin{equation}
\mathcal{D}[A]\rho=2A\rho A^{\dagger}-A^{\dagger}A\rho-\rho
A^{\dagger}A,
\end{equation}
for any given operator $A$.

Following similar procedures, substituting the value of $H_{\rm
ep}(t)$ in Eq.~(\ref{H-sb(t)}) into Eq.~(\ref{Electron-Phonon}) and
after some algebra, one obtains
\begin{eqnarray}
\mathcal{L}_{\rm ph}\rho_d^I(t)\!&\!=\!&\!\frac{{\gamma _2
}}{2}\mathcal{D}\left[ {\sigma_ +  } \right]\rho _d^I(t)  +
\frac{{\gamma _1 }}{2}\mathcal{D}\left[ {\sigma _ -  } \right]\rho
_d^I(t),\label{L-dis}
\end{eqnarray}
with
\begin{eqnarray}
\gamma _1\!&\!=\!&\!2\pi\left\{ {J\left( \varepsilon  \right)\left[
{n\left( \varepsilon  \right) + 1} \right] + J\left( { - \varepsilon
}
\right)n\left( { - \varepsilon } \right)} \right\},\nonumber\\
{\rm{   }}\gamma _2 \!&\!=\!&\!2\pi\left\{ {J\left( { - \varepsilon
} \right)\left[ {n\left( { - \varepsilon } \right) + 1} \right] +
J\left( \varepsilon  \right)n\left( \varepsilon  \right)} \right\},
\end{eqnarray}
where
\begin{eqnarray}
{J}\left( \omega  \right) = \sum\limits_q {\lambda _q^2 \delta
\left( {\omega  - \omega _q } \right)},
\end{eqnarray}
is the bath spectra density and
\begin{equation}
n\left( \varepsilon \right) = \frac{1}{{\exp \left( {\varepsilon
/k_B T} \right) - 1}},
\end{equation}
is the Bose-Einstein distribution.

With $\mathcal{L}_{\rm T}\rho_d^I(t)$ and $\mathcal{L}_{\rm
ph}\rho_d^I(t)$ given by Eq.~(\ref{L-T}) and Eq.~(\ref{L-dis}), the
master equation, Eq.~(\ref{A-ME-O}), for the reduced density matrix
of the DQD in the interaction picture is found to be
\begin{eqnarray}
\dot{\rho}_d^I(t) &=& \frac{{\Gamma _L }}{2}\mathcal{D}[ a_1 ] \rho
_d^I \bar f_l ( \omega_1)+ \frac{{\Gamma _L }}{2} \mathcal{D}[
a_1^{\dagger} ]\rho _d^I f_l ( \omega _1)\nonumber\\&&
+ \frac{{\Gamma _R }}{2}\mathcal{D}[ a_2 \Upsilon _r^{\dagger} ]
\rho_d^I \bar f_r \left( {\omega_2 } \right) + \frac{{\Gamma _R
}}{2} \mathcal{D}\left[ {a_2^{\dagger}  \Upsilon _r } \right]\rho
_d^I f_r \left( {\omega _2 } \right)\nonumber\\&&
+\frac{{\gamma _2}}{2}\mathcal{D}\left[ {\sigma _ +  } \right]\rho
_d^I + \frac{{\gamma _1 }}{2}\mathcal{D}\left[ {\sigma _ -  }
\right]\rho _d^I. \label{A-MEInter-O}
\end{eqnarray}
Next, we assume both a large bias voltage across the DQD (i.e.,
$\mu_L>\omega_1,~\omega_2>\mu_R$) and a very low temperature, so
that $f_l({\omega_1})=1,\,f_r(\omega_2)=0$.
After converting the resulting equation into the Schr\"{o}dinger
picture using the free evolution operator $e^{-iH_0t}$ or its
approximate in Eq. (\ref{U0}), we finally have
\begin{eqnarray}
\dot{\rho}_d(t)\!&=&\!-i[\,\frac{\varepsilon}{2}\sigma_z+{\Omega}\sigma_x,\,\rho_d(t)\,]
\nonumber\\&&
+ \frac{{\Gamma _L }}{2} \mathcal{D}[ a_1^{\dagger} ]\rho _d(t) +
\frac{{\Gamma _R }}{2}\mathcal{D}[ a_2 \Upsilon _r^{\dagger} ]
\rho_d(t) \nonumber\\&&
+\frac{{\gamma _2}}{2}\mathcal{D}\left[ {\sigma _ +  } \right]\rho
_d(t)  + \frac{{\gamma _1 }}{2}\mathcal{D}\left[ {\sigma _ -  }
\right]\rho _d(t),
\end{eqnarray}
which is just Eq.~(\ref{ME-O}), i.e., the master equation in the
occupation-state basis.

\section{Derivation of master equation in eigenstate basis}

This appendix gives further details on the derivation of the master
equation in the eigenstate basis given in Sec. IIB. Substituting
Eq.~(\ref{HT(t)-E}) into Eq.~(\ref{Tunneling}), and following
similar procedures in Sec. IIA, the dissipative part due to the
electric leads is evaluated to be
\begin{eqnarray}
&&\mathcal{L}_{\rm
T}\rho_d^I(t)=\frac{\Gamma_L}{2}\alpha^2\mathcal{D}[a_g^{\dagger}]\,\rho_d^I\,f_l(\omega_g)
+\frac{\Gamma_L}{2}\beta^2\mathcal{D}[a_e^{\dagger}]\,\rho_d^I\,f_l(\omega_e)
\nonumber\\&&
+\frac{\Gamma_L}{2}\alpha^2\mathcal{D}[a_g]\,\rho_d^I\,\bar
f_l(\omega_g)
+\frac{\Gamma_L}{2}\beta^2\mathcal{D}[a_e]\,\rho_d^I\,\bar
f_l(\omega_e) \nonumber\\&&
+\frac{\Gamma_R}{2}\beta^2\mathcal{D}[a_g^{\dagger}\Upsilon_r]\,\rho_d^I\,f_r(\omega_g)
+\frac{\Gamma_R}{2}\alpha^2\mathcal{D}[a_e^{\dagger}\Upsilon_r]\,\rho_d^I\,f_r(\omega_e)
\nonumber\\&&
+\frac{\Gamma_R}{2}\beta^2\mathcal{D}[a_g\Upsilon_r^{\dagger}]\,\rho_d^I\,\bar
f_r(\omega_g)
+\frac{\Gamma_R}{2}\alpha^2\mathcal{D}[a_e\Upsilon_r^{\dagger}]\,\rho_d^I\,\bar
f_r(\omega_e),\nonumber\\\label{B-L-T}
\end{eqnarray}
where $\alpha=\cos({\theta}/{2})~$ and $\beta=\sin(\theta/2)$. In
calculating Eq.~(\ref{B-L-T}), the fast oscillating terms
proportional to $e^{\pm i\Omega_0t}$ are neglected within the
rotating-wave approximation.
Similarly, from Eqs. (\ref{Hep(t)-E}) and (\ref{Electron-Phonon}),
the dissipative part due to the phonon bath reads
\begin{eqnarray}
&&\mathcal{L}_{\rm
ph}\rho_d^I(t)=\frac{\lambda_1}{2}\mathcal{D}[\sigma_-^{(e)}]\,\rho_d^I(t)
+\frac{\lambda_2}{2}\mathcal{D}[\sigma_+^{(e)}]\,\rho_d^I(t),\label{B-L-dis}
\nonumber\\&&
\end{eqnarray}
with the dissipation rates given by
\begin{eqnarray}
\lambda_1&=&2\pi{J}(\Omega_0)\cos^2{\theta}\,[\,n(\Omega_0)+1\,],\nonumber\\
\lambda_2&=&2\pi{J}(\Omega_0)\cos^2{\theta}\,n(\Omega_0),\,\label{A-dissipation
rate-E}
\end{eqnarray}
where
\begin{equation}
{J}(\Omega_0)=\sum_q\lambda_q^2\;\delta(\omega_q-\Omega_0),
\end{equation}
is the bath spectral density.

Substituting Eqs.~(\ref{B-L-T}) and (\ref{B-L-dis}) into
Eq.~(\ref{A-ME-O}), the master equation for the reduced density
matrix of the DQD in the interaction picture is
\begin{eqnarray}
&&\dot{\rho}_d^I(t)=\frac{\Gamma_L}{2}\alpha^2\mathcal{D}[a_g^{\dagger}]\,\rho_d^I\,f_l(\omega_g)
+\frac{\Gamma_L}{2}\beta^2\mathcal{D}[a_e^{\dagger}]\,\rho_d^I\,f_l(\omega_e)
\nonumber\\&&
+\frac{\Gamma_L}{2}\alpha^2\mathcal{D}[a_g]\,\rho_d^I\,\bar
f_l(\omega_g)
+\frac{\Gamma_L}{2}\beta^2\mathcal{D}[a_e]\,\rho_d^I\,\bar
f_l(\omega_e) \nonumber\\&&
+\frac{\Gamma_R}{2}\beta^2\mathcal{D}[a_g^{\dagger}\Upsilon_r]\,\rho_d^I\,f_r(\omega_g)
+\frac{\Gamma_R}{2}\alpha^2\mathcal{D}[a_e^{\dagger}\Upsilon_r]\,\rho_d^I\,f_r(\omega_e)
\nonumber\\&&
+\frac{\Gamma_R}{2}\beta^2\mathcal{D}[a_g\Upsilon_r^{\dagger}]\,\rho_d^I\,\bar
f_r(\omega_g)
+\frac{\Gamma_R}{2}\alpha^2\mathcal{D}[a_e\Upsilon_r^{\dagger}]\,\rho_d^I\,\bar
f_r(\omega_e),\nonumber\\&&
+\frac{\lambda_1}{2}\mathcal{D}[\sigma_-^{(e)}]\,\rho_d^I(t)
+\frac{\lambda_2}{2}\mathcal{D}[\sigma_+^{(e)}]\,\rho_d^I(t).\label{B-ME-E}
\end{eqnarray}
Here we also consider the case of both a large bias voltage across
the DQD (i.e., $\mu_L>\omega_g,~\omega_e>\mu_R$), and a very low
temperature, so that $
f_l(\omega_g)=f_l(\omega_e)=1,~~f_r(\omega_g)=f_r(\omega_e)=0.$
Converting Eq.~(\ref{B-ME-E}) into the Schr\"{o}dinger picture using
the free evolution operator $e^{-iH_ot}$ {\it without} needing
further approximation this time, the master equation of the reduced
density matrix of the DQD is given by
\begin{eqnarray}
\dot{\rho}_d(t)\!&=&\!-i[\,\frac{\Omega_0}{2}\sigma_z^{(e)},\,\rho_d(t)\,]
+\frac{\Gamma_L}{2}\alpha^2\mathcal{D}[a_g^{\dagger}]\,\rho_d(t)
\nonumber\\&&
+\frac{\Gamma_L}{2}\beta^2\mathcal{D}[a_e^{\dagger}]\,\rho_d(t)
+\frac{\Gamma_R}{2}\beta^2\mathcal{D}[a_g\Upsilon_r^{\dagger}]\,\rho_d(t)
\nonumber\\&&
+\frac{\Gamma_r}{2}\alpha^2\mathcal{D}[a_e\Upsilon_r^{\dagger}]\,\rho_d(t)
+\frac{\lambda_1}{2}\mathcal{D}[\sigma_-^{(e)}]\,\rho_d(t)
\nonumber\\&&
+\frac{\lambda_2}{2}\mathcal{D}[\sigma_+^{(e)}]\,\rho_d(t),
\end{eqnarray}
which is just Eq.~(\ref{ME-E}), i.e., the master equation in the
eigenstate basis. It should be emphasized that this master equation
is valid for arbitrary interdot coupling, in contrast to the master
equation in the occupation-state basis that is valid only for
small interdot coupling.


\end{document}